\documentclass[english,twosides,preprint,amsmath,nofootinbib]{revtex4-1}
\usepackage[T1]{fontenc}
\usepackage[latin9]{inputenc}
\usepackage{amsmath}
\usepackage{graphicx}
\usepackage{amssymb}
\usepackage{esint}

\makeatletter

\providecommand{\tabularnewline}{\\}

\@ifundefined{textcolor}{}
{%
 \definecolor{BLACK}{gray}{0}
 \definecolor{WHITE}{gray}{1}
 \definecolor{RED}{rgb}{1,0,0}
 \definecolor{GREEN}{rgb}{0,1,0}
 \definecolor{BLUE}{rgb}{0,0,1}
 \definecolor{CYAN}{cmyk}{1,0,0,0}
 \definecolor{MAGENTA}{cmyk}{0,1,0,0}
 \definecolor{YELLOW}{cmyk}{0,0,1,0}
 }

\@ifundefined{definecolor}{\@ifundefined{definecolor}{\@ifundefined{definecolor}{\@ifundefined{definecolor}{\@ifundefined{definecolor}{\@ifundefined{definecolor}{\@ifundefined{definecolor}
 {\usepackage{color}}{}
}{}}{}}{}}{}}{}}{}\usepackage[dvipdfm,bookmarks=true]{hyperref}

\makeatother

\usepackage{babel}

\makeatother

\usepackage{babel}

\makeatother

\usepackage{babel}

\makeatother

\usepackage{babel}

\makeatother

\usepackage{babel}

\makeatother

\usepackage{babel}

\makeatother

\usepackage{babel}

\makeatother

\usepackage{babel}

\makeatother

\usepackage{babel}

\makeatother

\usepackage{babel}

\begin{document}

\title{{\LARGE Gamma-rays from Nearby Clusters: Constraints on Selected
Decaying Dark Matter Models}}

\date{\today}

\author{Jiwei Ke}

\email{kejiwei1985@gmail.com}

\author{Mingxing Luo}

\email{luo@zimp.zju.edu.cn}

\author{Liucheng Wang}

\email{liuchengwang@zimp.zju.edu.cn}

\author{Guohuai Zhu}

\email{zhugh@zju.edu.cn;(Corresponding author)}

\affiliation{Zhejiang Institute of Modern Physics, Department of Physics, Zhejiang
University, Hangzhou, Zhejiang 310027, P.R.China}
\begin{abstract}
Recently, the Fermi-LAT collaboration reported upper limits on the
GeV gamma-ray flux from nearby clusters of galaxies. Motivated by
these limits, we study corresponding constraints on gamma-ray
emissions from two specific decaying dark matter models, one via grand
unification scale suppressed operators and the other via R-parity
violating operators. Both can account for the PAMELA and Fermi-LAT
excesses of $e^{\pm}$. For GUT decaying dark matter, the gamma-rays
from the M49 and Fornax clusters, with energy in the range of $1$
to $10$ GeV, lead to the most stringent constraints to date. As a
result, this dark matter is disfavored with conventional model of
$e^{\pm}$ background. In addition, it is likely that some tension exists between
the Fermi-LAT $e^\pm$ excess and the gamma-ray constraints for any decaying dark matter model, provided conventional model
of $e^{\pm}$ background is adopted. Nevertheless, the GUT decaying dark matter can still solely account for
the PAMELA positron fraction excess without violating the gamma-ray
constraints. For the gravitino dark matter model with R-parity violation,
cluster observations do not give tight constraints. This is because
a different $e^{\pm}$ background has been adopted which leads to relatively
light dark matter mass around 200 GeV.
\end{abstract}
\maketitle

\section{Introduction}

It is now believed that the dominant matter in the universe should
be non-baryonic dark matter (DM) instead of visible ones. In addition,
DM should not be composed of any known Standard Model (SM) particles.
There are three experimental avenues to prove the existence of DM:
direct detection, indirect detection and collider searches. In this
paper, we will focus on the indirect detection of DM through gamma-rays
and electrons/positrons in cosmic-rays. The Pamela collaboration has
reported significant excess in the positron fraction between $1$
and $100$ GeV \cite{Adriani:2008zr}. Later on the Fermi-LAT collaboration
also observed a clear feature in the spectrum of electrons and positrons
from $20$ GeV to $1$ TeV \cite{Abdo:2009zk}, which is harder than
predictions of conventional models of cosmic ray propagation.
These excesses of electrons and positrons could be attributed by the
annihilation or decay of DM. If so, one might also detect gamma-rays
from these DM annihilation/decay.

Recently, the Fermi-LAT collaboration has reported measurements of the GeV gamma-ray
from nearby clusters of galaxies \cite{Ackermann:2010qj}.
In an observation period of 18-month, there was no direct observational
evidence. The upper limits on gamma-ray flux are given with 95\% confidence-level.
As clusters of galaxies are DM dominated and DM annihilation/decay
would almost inevitably emit gamma-rays, these upper limits may lead
to stringent constraints on DM model parameters. Notice that based
on observations of the first 11-month of Fermi-LAT, there already existed
some discussions about DM models along this way \cite{Ackermann:2010rg,Yuan:2010gn,Dugger:2010ys}.

To interpret the PAMELA and Fermi-LAT excesses in terms of DM annihilation,
a large boost factor of order $100-1000$ is required for the theory
to be consistent with the relic abundance measured by the WMAP \cite{Bergstrom:2008gr,Cholis:2008hb,Cirelli:2008pk}.
Moreover, gamma-ray fluxes produced by the DM annihilation are anticipated
to be large at locations with very high density of DM, because they
are proportional to the local DM density squared. There are plenty
of experimental measurements of gamma-rays from inside/outside the
Galactic halo. Accordingly, the annihilating DM scenarios have been
strongly constrained by the observations \cite{Nardi:2008ix,Bertone:2008xr,Cirelli:2009dv,Papucci:2009gd,Ackermann:2010rg,Yuan:2010gn}.
So it is not easy to build models of annihilating DM at TeV scale
satisfying all constraints.\footnote{There are certainly ways out of this problem, see for example a recent
attempt in \cite{Liu:2011aa}%
}

Decaying DM offers an alternative explanation to the $e^{\pm}$ excesses.
In decaying DM scenarios, constraints from gamma-ray fluxes are relatively
easier to be satisfied since the gamma-ray fluxes are linearly proportional
to the local DM density. However to fit the PAMELA and Fermi-LAT excesses,
a lifetime of the order $10^{26}$'s for DM is needed, which is even
much longer than that of the Universe. From the perspective of particle
physics, there are several frameworks which can naturally provide such a long
lifetime. For instance, DM may decay via operators suppressed
by the grand unification theory (GUT) scale $10^{16}$ GeV, which
makes it sufficiently long-living. DM may also decay via a very weak
R-parity violation process: R-parity conservation makes the lightest
supersymmetric particle (LSP) stable to be DM candidate while R-parity
violating operators make DM decay sufficiently slow. There are of
course other possibilities, for example DM may decay via instanton-induced
operators \cite{Carone:2010ha}. Recently it has also been shown that
such a lifetime can arise naturally from the goldstino decay \cite{Cheng:2010mw}.

For decaying DM with $\mu^{+}\mu^{-}$ and $b\bar{b}$ final states,
the impact of the Fermi 11-month observation of gamma-rays from nearby clusters
has been discussed in \cite{Dugger:2010ys}. It was found that these gamma-rays
from clusters considerably improves previous constraints on the lifetime
and mass of the DM. Especially, the Fornax cluster provides the strongest
constraint to date. This motivated us to investigate similar constraints
on some (arguably) theoretically better-motivated models of decaying
DM. Specifically, we will focus on two different decaying DM scenarios
mentioned above: decaying via GUT scale suppressed operators or decaying
via R-parity violating operators. As shown before, each scenario could
give a reasonable fit to the $e^{\pm}$ excesses. In addition, both
scenarios are consistent with measurements of Galactic and extragalactic
gamma-rays. We expect that the gamma-rays from nearby clusters should
lead to more stringent constraints than before.

This paper is organized as follows. We discuss the gamma-ray fluxes
from galaxy clusters in general in Section II. In Section III, DM
gamma-ray signals from nearby clusters are discussed in a GUT framework.
Section IV is devoted to the study of a decaying DM scenario with
R-parity violating operators. We conclude with a summary in section
V.

\section{Gamma-rays from clusters in decaying DM scenarios}

Nearby clusters are expected to be very interesting targets for DM
indirect detection \cite{Jeltema:2008vu}. They are supposed to be
highly DM dominated and isolated at high galactic latitudes. Thus
high signal-to-noise ratios are anticipated for gamma-ray observations
targeting nearby clusters. Certainly there may have other sources
in clusters that can emit gamma-rays, besides DM annihilation/decay.
Nevertheless the gamma-ray observations of nearby clusters can give
upper limits to certain DM model parameters. Along this way, the EGRET
gamma-ray measurements have been used to constrain DM annihilation
models in \cite{Pinzke:2009cp}. Last year Fermi-LAT reported
interesting measurements of gamma-rays from galaxy clusters of a 11-month
observation period, which motivated new research efforts on DM annihilation
\cite{Ackermann:2010rg,Yuan:2010gn,Cuesta:2010ex} and DM decay \cite{Cuesta:2010ex,Dugger:2010ys}.
Recently,the Fermi-LAT collaboration have updated their results based
on an observation of 18-month \cite{Ackermann:2010qj}, which we will
investigate in the following.

In general, the gamma-ray flux coming from a galaxy cluster can be
written as \begin{equation}
\Phi(E_{\gamma})=J(\psi)\times W(E_{\gamma}),\end{equation}
 where $J(\psi)$ and $W(E_{\gamma})$ encode the astrophysical information
and particle information, respectively. Along the direction of a cluster,
$J(\psi)$ is defined as \begin{equation}
J(\psi)=\int_{\triangle\Omega}d\Omega\int dl(\psi)\rho_{DM}(l).\label{eq:J}\end{equation}
 Here the solid angle $\triangle\Omega$ corresponds to a radius of
ten degrees, since the Fermi-LAT Collaboration selected all gamma-rays
within a $10^{\circ}$ radius around the direction of each galaxy
cluster in their sample \cite{Ackermann:2010qj}. The parameter $l(\psi)$
is integrated along a line of sight within each cluster and $\rho_{DM}(l)$
is the DM mass density. Since $J(\psi)$ depends only on the integration
over $\rho_{DM}(l)$, unlike the case of annihilating DM, the gamma-ray
flux here is rather insensitive to the DM density profile.

In this paper, we will consider six galaxy clusters reported by
Fermi-LAT. They are the NGC4636, M49, Fornax, Centaurus, AWM7 and
Coma clusters. All of them are DM dominated and have very low gamma-ray
backgrounds, as they are isolated at high galactic latitudes. The
upper limits of their gamma-ray fluxes may give severe constraints
on decaying DM models. More information about these galaxy clusters
are listed in Table \ref{tab:6 clusters}.

\begin{table}
\begin{tabular}{|c|c|c|c|c|c|c|c|}
\hline
Name  & z  & R.A.  & Dec.  & $R_{200}$  & $M_{200}$  & $R_{500}$  & $M_{500}$\tabularnewline
\hline
\hline
NGC4636  & 0.0031  & 12h43m  & $2^{\circ}41'$  & 0.85  & 0.35  & 0.53  & 0.22\tabularnewline
\hline
M49  & 0.0033  & 12h30m  & $8^{\circ}00'$  & 1.04  & 0.65  & 0.66  & 0.41\tabularnewline
\hline
Fornax  & 0.0046  & 3h39m  & $-35^{\circ}27'$  & 1.35  & 1.42  & 0.84  & 0.87\tabularnewline
\hline
Centaurus  & 0.0114  & 12h49m  & $-41^{\circ}18'$  & 1.87  & 3.74  & 1.18  & 2.33\tabularnewline
\hline
AWM7  & 0.0172  & 2h55m  & $41^{\circ}35'$  & 2.19  & 6.08  & 1.38  & 3.79\tabularnewline
\hline
Coma  & 0.0231  & 13h00m  & $27^{\circ}59'$  & 3.22  & 19.38  & 2.03  & 11.99\tabularnewline
\hline
\end{tabular}\caption{\label{tab:6 clusters} Cluster Information. Redshifts and equatorial
coordinates of clusters can be found in the NASA/IPAC Extragalactic
Database. Virial radius $R_{200}$ (a radius within which the mean
cluster density is 200 times larger than the critical density $\rho_{c}$),
$R_{500}$, and corresponding viral masses $M_{200}$, $M_{500}$
are adopted from \cite{Reiprich:2001zv}. $R_{200}$ and $R_{500}$
are in the unit of $(2h)^{-1}$ Mpc, where h is the present day normalized
Hubble expansion rate. $M_{200}$ and $M_{500}$ are in the unit of
$(2h)^{-1}$$10^{14}M_{\odot}$, where $M_{\odot}$ is the solar mass.}

\end{table}

Since the Fermi-LAT analysis is done within a $10^{\circ}$ radius
surrounding the cluster center, the corresponding spatial extent is
even larger than the cluster virial radius $R_{200}$. So we integrate
over the whole cluster halo in Eq (\ref{eq:J}). Moreover, the cluster
radius $R_{200}$ is much smaller than its luminosity distance $D$.
Approximately as a point source, we get\begin{equation}
J(\psi)\simeq\frac{1}{D^{2}}\int dV\rho_{DM}(l)=\frac{M_{200}}{D^{2}}.\end{equation}
So for each cluster, $J(\psi)$ is approximately independent of specific
DM mass density assumption $\rho_{DM}(l)$. Using data in Table \ref{tab:6 clusters},
we find that the Fornax cluster has the largest ratio $M_{200}/D^{2}$,
which should have the brightest gamma-ray emission from decaying DM.

Now we turn to the particle physics factor $W(E_{\gamma})$, which
contains the mechanism to produce photons in decaying DM scenarios.
The gamma-ray flux comes from final state radiations (FSR) and the
inverse Compton scattering (ICS).

(1) FSR: Inevitably the bremsstrahlung of $e^{\pm}$ interacting with
interstellar and/or intracluster gas leads to the emission of energetic
photon. In addition, if $\tau$ lepton, for example, exists in the
decay process, $\tau\rightarrow\pi^{0}\rightarrow\gamma+\gamma$ would
also produce gamma-rays. FSR is quite model-dependent and all decay
channels involving photons should be taken into account. Notice that
photons propagate almost freely, the FSR flux of photons from a cluster
is given by \begin{equation}
\Phi(E_{\gamma})=\frac{M_{200}}{4\pi D^{2}}\underset{i}{\sum}\frac{\Gamma_{i}^{DM}}{M^{DM}}\frac{dN_{i}^{DM}}{dE_{\gamma}}.\end{equation}
 Here the summation is over all possible decay channels which may
produce gamma-rays. $\Gamma_{i}^{DM}$ is the decay width of a specific
channel and $M^{DM}$ is the DM mass. $dN_{i}^{DM}/dE_{\gamma}$ is
the photon spectrum per DM decay via a specific channel. PYTHIA package
\cite{Sjostrand:2006za} has been used in our calculation to obtain
these spectra numerically.

(2) ICS: The ICS radiation is produced when the energetic $e^{\pm}$
scatter on the interstellar radiation field (ISRF) in galaxy clusters.
A pedagogical review of ICS $W(E_{\gamma})$ is provided in \cite{Blumenthal:1970gc}.
We calculate it semi-analytically, following Refs. \cite{Cirelli:2009vg,Ibarra:2009dr,Ibarra:2009nw,Ishiwata:2009dk,Luo:2009xd,Zhang:2009kp}.
Noticing that the CMB component is dominant in the ISRF in each cluster.
Combining $J(\psi)$ and $W(E_{\gamma})$, we obtain the ICS flux
of photons from a cluster \begin{equation}
\Phi(E_{\gamma})=\frac{M_{200}}{4\pi D^{2}}\underset{i}{\sum}\frac{\Gamma_{i}^{DM}}{M^{DM}}\int d\epsilon f_{CMB}(\epsilon)\int dE_{e}\frac{d\sigma^{ICS}(E_{e},\epsilon)}{dE_{\gamma}}\frac{dN_{i}^{DM}(E_{e})}{dE_{e}}.\end{equation}
 Here the summation is over all possible decay channels with electron
and/or positron in the final states. The number density of CMB background
$f_{CMB}(\epsilon)$ should be a blackbody-like spectrum \cite{Cirelli:2009vg}
\begin{equation}
f_{CMB}(\epsilon)=\frac{\epsilon^{2}}{\pi^{2}}\frac{1}{e^{\epsilon/T}-1}\end{equation}
 with $T=2.753$ K. The initial electron/positron spectrum may be
solved from the diffusion-loss equation. Approximately, \begin{equation}
\frac{dN_{i}^{DM}(E_{e})}{dE_{e}}=\frac{1}{B(E_{e})}\int_{E_{e}}^{M_{DM}}dE'\frac{dN_{i}^{DM}}{dE'},\end{equation}
 where $B(E_{e})=E^{2}/(GeV\cdot\tau_{E})$ is the effective energy
loss coefficient with $\tau_{E}=4\times10^{16}$s. The Compton cross
section is given by the Klein-Nishina formula \begin{equation}
\frac{d\sigma^{ICS}(E_{e},\epsilon)}{dE_{\gamma}}=\frac{3\sigma_{T}}{4\gamma_{e}^{2}\epsilon}\left(2q\ln q+1+q-2q^{2}+\frac{(q\Gamma)^{2}}{2(1+q\Gamma)}(1-q)\right)\end{equation}
 where \begin{equation}
q=\frac{E_{\gamma}}{\Gamma(E_{e}-E_{\gamma})},~\Gamma=\frac{4\gamma_{e}\epsilon}{m_{e}},~\gamma_{e}=\frac{E_{e}}{m_{e}}~.\end{equation}
 $m_{e}$ denotes the mass of electron and $\sigma_{T}=0.67$ barn
is the Compton scattering cross section in the Thomson limit. For
energy integration, $\epsilon\leq E_{\gamma}\leq[(1/E_{e}+1/(4\gamma_{e}^{2}\epsilon)]^{-1}$
is required by kinematics.

\section{Decaying dark matter in grand unification theory}

\begin{figure}[tb]
 \includegraphics[scale=1.0]{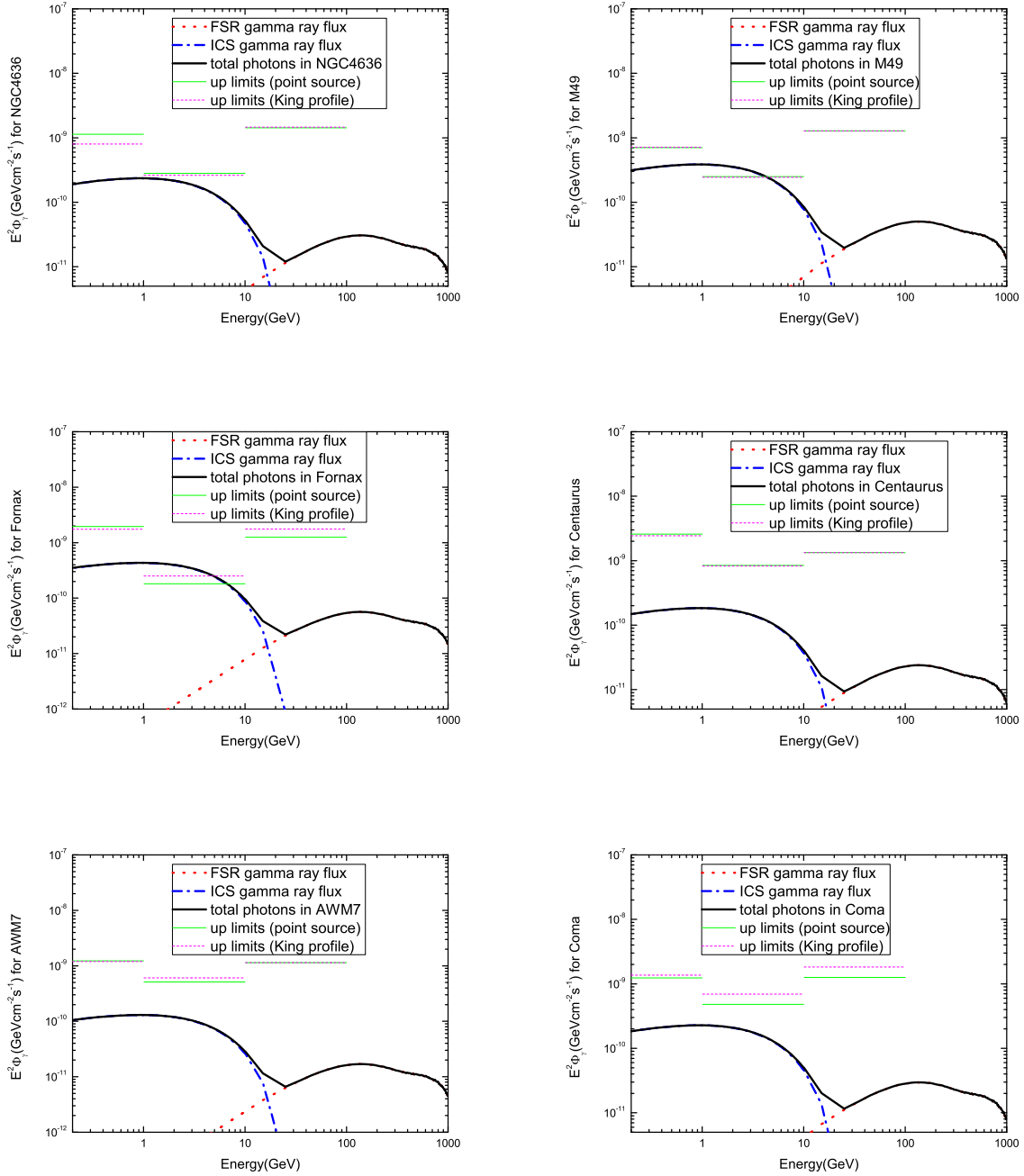}
\caption{\label{fig:GUT}In a GUT DM model, gamma-ray spectra are shown for
NGC4636 (top left), M49 (top right), Fornax (middle left), Centaurus
(middle right), AWM7 (bottom left) and Coma (bottom right) clusters.
Experimental upper limits are from the Fermi-LAT collaboration \cite{Ackermann:2010qj}.}
\end{figure}

The possibility that the Pamela positron fraction excess may be accounted for by the GUT suppressed operators was first discussed in \cite{Chen:2008md}. Since then, many decaying DM models in the framework of GUT have been proposed \cite{Arvanitaki:2008hq,Arvanitaki:2009yb,Luo:2009xd,Ruderman:2009ta,Ruderman:2009tj,Kadastik:2009cu,Kyae:2009gm,Haba:2010ag,Gao:2010pg}.
To quantitatively study the possible gamma-ray signals from clusters,
we choose one of the DM models as a template \cite{Luo:2009xd}. First
we briefly review the main features of this model. An SU(5) singlet
$S$ is introduced as the DM candidate. For $S$ to be stable, a discrete
$Z_{2}$ symmetry is assumed. $S$ couples to the MSSM particles via
dimension six operators suppressed by GUT scale $M_{GUT}\sim10^{16}$
GeV \begin{equation}
\frac{S^{+}S\overline{5}^{+}\overline{5}}{M_{GUT}^{2}},~~~~~~\frac{S^{+}STr(10^{+}10)}{M_{GUT}^{2}}~.\label{eq:dimension-6 operators}\end{equation}
 Here $S=\tilde{s}+\sqrt{2}\theta s+\theta^{2}F_{s}$ is a standard
chiral superfield. $\overline{5}$ is the anti-fundamental representation
of SU(5) and $10$ is the antisymmetric tensor representation.

For DM decay, we further assume the $Z_{2}$ symmetry to be spontaneously
broken. The scalar field of $S$ develops a vacuum expectation value
(VEV), making both fields $(\widetilde{s},\: s)$ in $S$ decay via
the operators of Eq.(\ref{eq:dimension-6 operators}). In addition,
the squark masses are assumed to be so heavy that DM decays dominantly
into sleptons. Assuming the DM is mainly composed of scalar $\tilde{s}$,
the operators of Eq.(\ref{eq:dimension-6 operators}) can be expanded
in terms of component fields as \begin{equation}
\underset{\widetilde{l}}{\sum}\frac{-1}{M_{GUT}^{2}}<\widetilde{s}>\widetilde{s}^{*}(\widetilde{l_{L}}^{*}\square\widetilde{l_{L}}+\widetilde{l_{R}}\square\widetilde{l_{R}}^{*}).\end{equation}
 Here $\widetilde{l}$ denotes the sleptons $\widetilde{e},~\widetilde{\mu}$
and $\widetilde{\tau}$. With R-parity conservation, the slepton would
decay to the LSP and lepton subsequently. $e^{\pm}$ and $\gamma$
can be produced through the following cascade decay chains: (1) selectron
chain $\widetilde{s}\rightarrow\widetilde{e}\rightarrow e/\gamma$;
(2) smuon chain $\widetilde{s}\rightarrow\widetilde{\mu}\rightarrow\mu\rightarrow e/\gamma$
and (3) stau chain $\widetilde{s}\rightarrow\widetilde{\tau}\rightarrow\tau\rightarrow e/\gamma$.
Numerically, PYTHIA package \cite{Sjostrand:2006za} is used to get
the spectra of $e^{\pm}$ and $\gamma$.

For interstellar $e^{\pm}$ background fluxes, we adopt {}``model
0'' presented by the Fermi-LAT \cite{Abdo:2009zk}. Reasonable fits
to PAMELA $e^{+}/(e^{+}+e^{-})$ fraction and Fermi-LAT $e^{+}+e^{-}$
spectrum can be obtained \cite{Luo:2009xd} with the illustrative
parameter set: DM mass $M_{\widetilde{s}}=6.5$ TeV, $M_{GUT}=10^{16}$
GeV, the VEV $<\widetilde{s}>=20$ TeV, selectron mass $M_{\widetilde{e}}=380$
GeV, smuon mass $M_{\tilde{\mu}}=370$ GeV, stau mass $M_{\widetilde{\tau}}=330$
GeV and LSP particle mass$M_{LSP}=300$ GeV. Moreover, gamma-rays
emitted from DM decay are consistent with the Fermi-LAT data in the
region $0^{\circ}\leq l\leq360^{\circ},\:10^{\circ}\leq|b|\leq20^{\circ}$.

\begin{figure}[tb]
 \includegraphics[scale=1.0]{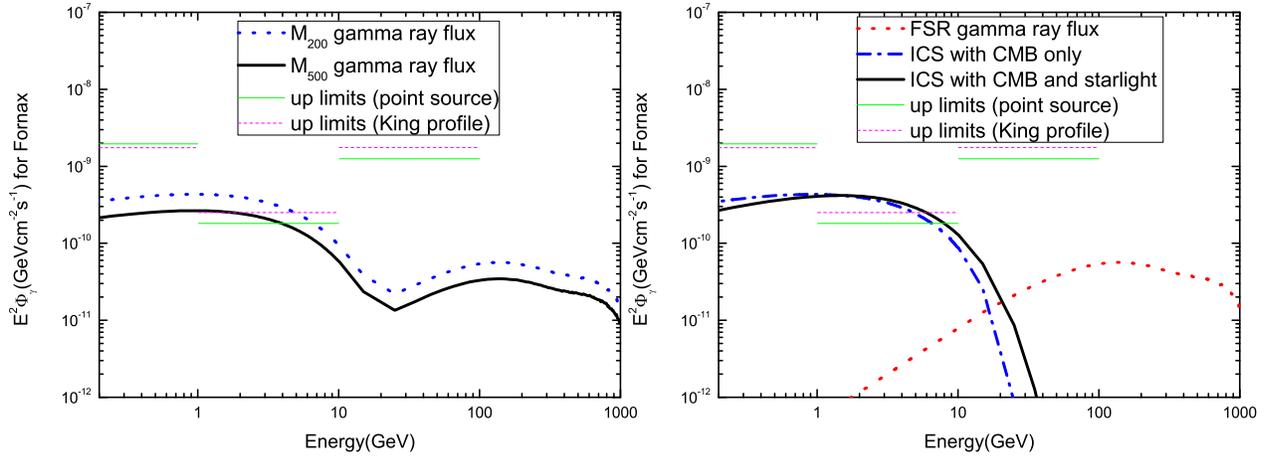}
\caption{\label{fig:Fornax}Left: the uncertainty of total DM mass in Fornax
cluster, the thick solid curve represents the gamma-ray flux with
virial mass $M_{500}$. Right: estimating the starlight in the Fornax
cluster in a naive way. The corresponding ICS flux of photons is shown
as a thick solid curve. To contrast, the standard gamma-ray fluxes
are reproduced from Fig \ref{fig:GUT}.}
\end{figure}

With the same parameter set, the gamma-ray signals coming from six
nearby clusters are compared with the upper limits measured recently
by the Fermi-LAT collaboration \cite{Ackermann:2010qj}, as shown
in Fig. \ref{fig:GUT}. The expected signals and experimental upper
limits are comparable in magnitude for all six clusters. Therefore
clusters are good targets to constrain or even falsify DM models.
As we can see, this GUT DM model already predicts too much gamma-rays
from the M49 and Fornax clusters in the energy range $1-10$ GeV,
which are in disagreement with the Fermi-LAT observations. Coincidentally,
the $1-10$ GeV energy band is also the most sensitive range of the
Fermi-LAT detector. In this energy range the gamma rays come mainly
from the ICS process, as can be seen from Fig. \ref{fig:GUT}. As
the CMB photons are well measured, ICS process is largely determined
by the spectra of initial electrons and positrons.
Notice that for all decaying DM models which could account for the PAMELA and Fermi-LAT $e^\pm$ excesses,
the produced spectra of electrons and positrons should be more or less the same provided the same $e^\pm$ background is used.
In this sense, the ICS flux of photons is model-independent to a large extent.
So it is likely that, for any decaying dark matter model, some tension exists between
the Fermi-LAT $e^\pm$ excess and the gamma-ray constraints from nearby clusters if conventional model
of $e^{\pm}$ background is adopted.

\begin{figure}[tb]
 \includegraphics[scale=1.0]{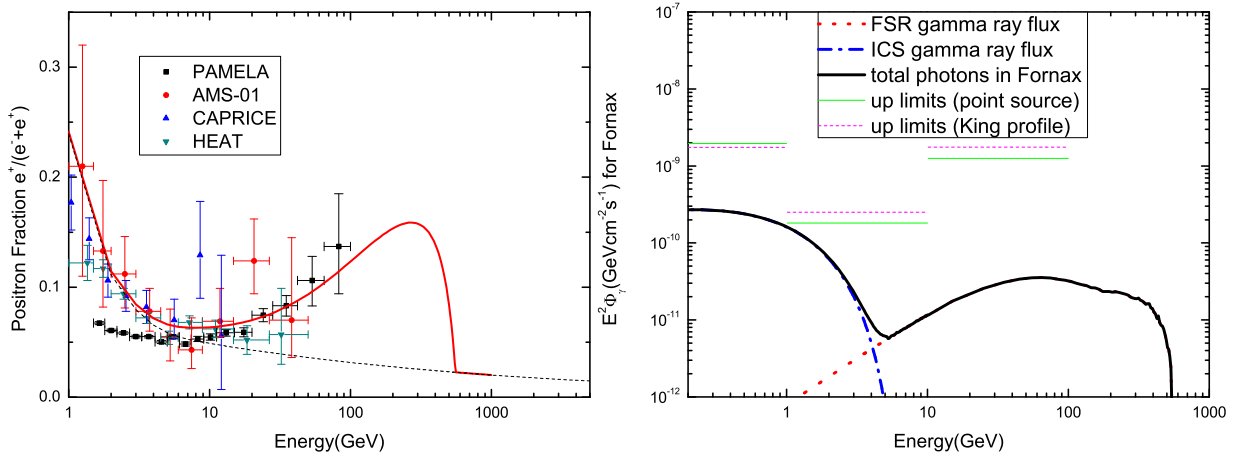}
\caption{\label{fig:New fit}New parameter set in a GUT decaying DM model.
Left: the predicted positron fraction is consistent with experimental
data. The solid line shows the fit and the dash line denotes the background.
Right: DM induced gamma-ray fluxes in the Fornax cluster, which are
below the Fermi-LAT upper limits.}
\end{figure}

We now address the theoretical uncertainties of gamma-ray
signals. As we have taken the point source approximation in Section
II, the Fornax cluster gives a severer constraint than the M49 cluster
does. So let's take the Fornax cluster as an example. One source of
uncertainties is related to the total DM mass in the Fornax cluster.
For example, if we choose virial radius $R_{500}$ instead of $R_{200}$,
the correspondingly total DM mass is $M_{500}$, which is less than
$M_{200}$. Even so our model still overproduces gamma-ray flux, as
shown on the left of Fig \ref{fig:Fornax}. Another source of uncertainties
is that only the CMB photons are considered in the ICS process. In
principle, the energetic $e^{\pm}$ may also scatter on the starlight
in the Fornax cluster, which may lead to a harder gamma-ray spectrum.
However, it is rather difficult to quantitatively estimate the effect
of starlight in the Fornax cluster, though qualitatively one might
expect that this effect should not be too large for such a DM dominant
system. In order to have a feeling on the potential starlight effect,
we relate naively the starlight of Fornax cluster to that of our Milky
Way by a dimensional analysis \begin{equation}
\frac{f_{Fornax}}{f_{Milky}}=\frac{L_{Fornax}/R_{Fornax}^{3}}{L_{Milky}/R_{Milky}^{3}}~,\end{equation}
 where $f$ is the starlight photon density. The total luminosity
$L$ in the Milky Way is about $10^{44}$ erg/s while the Fornax luminosity
is taken from \cite{Reiprich:2001zv}. Choosing the density $f_{Milky}$
in the range $0^{\circ}\leq l\leq360^{\circ},\:10^{\circ}\leq|b|\leq20^{\circ}$
as a typical value, the starlight density $f_{Fornax}$ can be estimated.
The resulting gamma-ray spectrum is shown in the right of Fig \ref{fig:Fornax},
which indicates that the impact of starlight on the gamma-rays is
indeed rather limited. Though the ICS spectrum does become harder
as expected, it still violates the Fermi-LAT constraints in the energy
range $1-10$ GeV.

The way out of this problem may lie in the change of cosmic ray background.
For example, if the Fermi-LAT electrons excess is due to unidentified
astrophysical sources instead of DM decay, a new parameter set can
be chosen to fit the Pamela excess: $M^{DM}=3$ TeV, $M_{GUT}=10^{16}$
GeV, $<\widetilde{s}>=11$ TeV, $M_{\widetilde{e}}=380$ GeV, $M_{\tilde{\mu}}=370$
GeV, $M_{\widetilde{\tau}}=330$ GeV and $M_{LSP}=300$ GeV. In this
case the DM mass is smaller by a factor of two, correspondingly the
DM induced gamma-ray spectrum becomes softer. Then the gamma-ray flux
upper limits from the Fornax cluster can be satisfied, as shown in
Fig. \ref{fig:New fit}. It may also be possible that the Fermi-LAT
electrons excess is accounted for partly by the change of electrons
background and partly by the DM contribution. We will discuss this
possibility in the next section, though in a different decaying DM
model with R-parity violation.

\section{Decaying dark matter with R-parity violation}

\begin{figure}[tb]
 \includegraphics[scale=1.0]{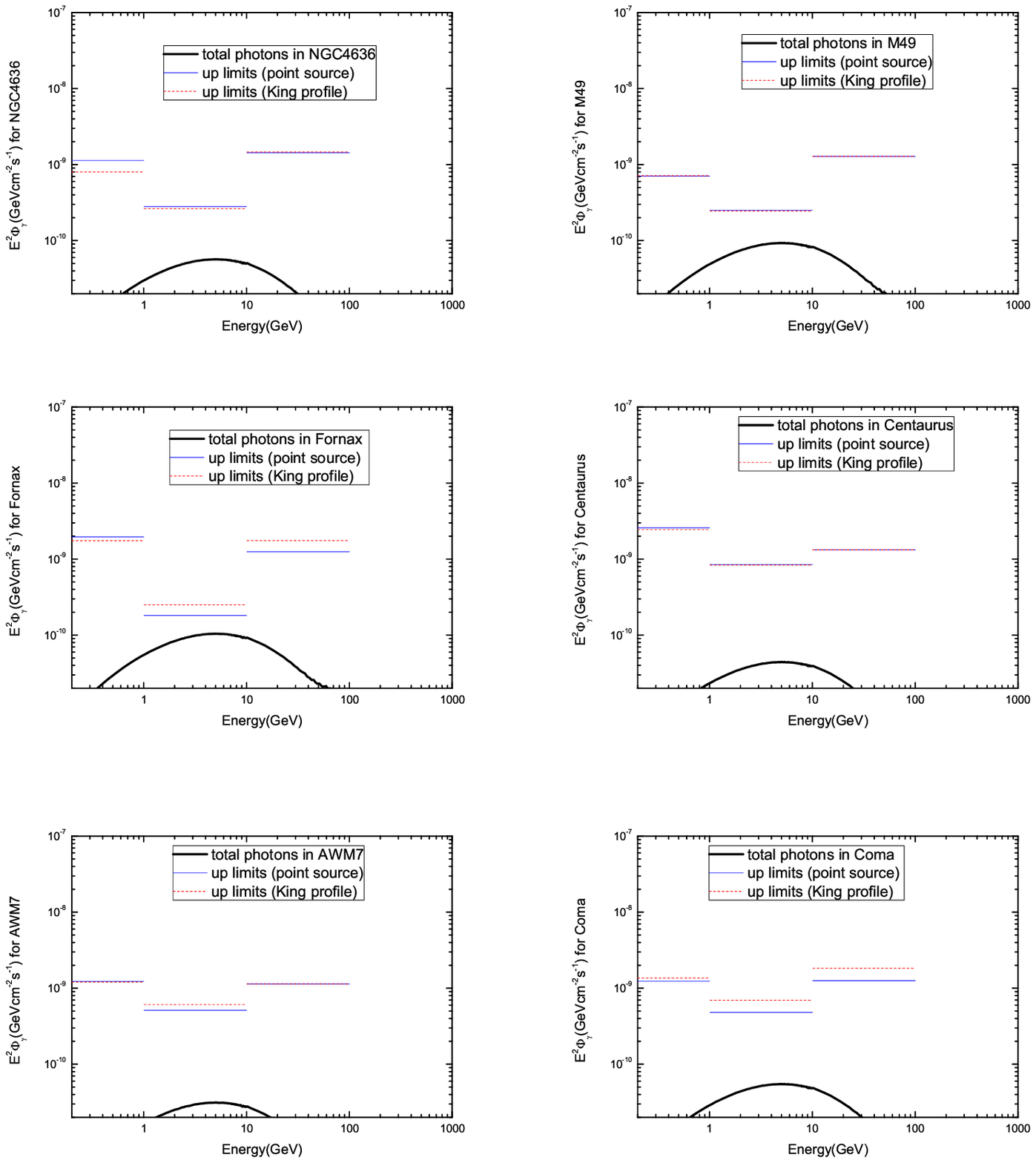}
\caption{\label{fig:RPV}For decaying DM with R-parity violation, gamma-ray
spectra from NGC4636 (top left), M49 (top right), Fornax (middle left),
Centaurus (middle right), AWM7 (bottom left) and Coma (bottom right)
clusters are shown. Experimental upper limits are from the Fermi-LAT
collaboration \cite{Ackermann:2010qj}. The main contribution to GeV
gamma-ray signals comes from FSR process in this model}
\end{figure}

In R-parity violating SUSY models, the gravitino is an interesting DM candidate as primordial nucleosynthesis, thermal leptogenesis and
gravitino DM are naturally consistent \cite {Buchmuller:2007ui}.
Motivated by the PAMELA and Fermi-LAT anomalies,
considerable efforts have been devoted to study decaying DM in the
framework of SUSY models with R-parity violation \cite{Ishiwata:2008cu,Ishiwata:2010am,Ishiwata:2009vx,Shirai:2009fq,Chen:2009ew,Hamaguchi:2009sz,Vertongen:2011mu,
Choi:2010jt,Bomark:2009zm,Buchmuller:2009xv}.

Phenomenologically, the trilinear operator $L_i L_j \bar{E}_3$  with the gravitino mass heavier than $1.5$ TeV could fit the Pamela and Fermi-LAT
data \cite{Bomark:2009zm}, where $L/E$ are the left-handed lepton doublet/singlet
superfields. Since both the GUT decaying DM model and the above gravitino DM model
involve multibody decay processes, the produced spectra of electrons and positrons in both models should be
more or less the same to explain the Pamela and Fermi-LAT excesses.
But as discussed in the previous section, gamma rays in the energy range $1-
10$ GeV come mainly from the ICS process which is largely determined
by electrons and positrons produced from the DM decays.
Therefore it is likely that, similar to the case of GUT decaying DM, the
$LL\bar{E}$-type gravitino model would also predict too much gamma-rays
from the Fornax clusters in the energy range $1-10$ GeV as compared with the
Fermi-LAT observations.

Here, we will focus instead on a decaying gravitino ($\psi_{\mu}$) DM model
\cite{Ishiwata:2008cu}, where R-parity violating operators are introduced
in the soft supersymmetry breaking lagrangian as \begin{equation}
{\cal L}_{RPV}=B_{i}\tilde{L_{i}}H_{u}+M_{\tilde{L_{i}},H_{d}}^{2}\tilde{L_{i}}H_{d}^{*}+H.C.\end{equation}
 Here $H_{u}$ and $H_{d}$ are up-type and down-type Higgs doublet,
respectively. $\tilde{L_{i}}$ is left-handed slepton doublet, the
index $i$=1, 2, 3 denotes the sleptons $\widetilde{e},~\widetilde{\mu}$
and $\widetilde{\tau}$, respectively. R-parity and the lepton number
are spontaneously breaking if the left-handed sneutrino field $\tilde{\nu}_{i}$
gets a VEV $<\tilde{\nu}_{i}>$. Then the DM particle $\psi_{\mu}$
will no longer be absolutely stable. By integrating out heavier fields,
one can obtain the effective lagrangian governing the $\psi_{\mu}$
decay, as discussed in \cite{Ishiwata:2008cu}. A proper small $<\tilde{\nu}_{i}>$
would lead to the lifetime of DM around $10^{26}$s. There are various
cascade decay chains which may produce energetic electrons, positrons
and gamma-rays: (1) $Z$ channel $\psi_{\mu}\rightarrow Z+\nu\rightarrow e/\gamma$;
(2) $W$ channel $\psi_{\mu}\rightarrow W+l\rightarrow e/\gamma$;
(3) Higgs channel $\psi_{\mu}\rightarrow h+\nu\rightarrow e/\gamma$
and (4) Monochromatic channel $\psi_{\mu}\rightarrow\gamma+\nu$ (no
$e^{\pm}$ final state). The relative importance of different decay
chains depends very sensitively on the DM mass. Again PYTHIA package
\cite{Sjostrand:2006za} has been used in order to describe the hadronization
process and get the energy spectra of $e^{\pm}$ and $\gamma$.

Phenomenology of this model has been discussed in detail in \cite{Ishiwata:2010am}.
Assuming $B_{1}\gg B_{2}$, $B_{3}$, the DM $\psi_{\mu}$ will decay
mainly into the first generation leptons plus $\gamma$, W, Z and
Higgs bosons. Assuming large Higgsino-mass limit and gaugino mass
unification, the parameters can be chosen as: DM mass $M_{\psi_{\mu}}$=
200 GeV, DM lifetime $\tau_{\psi_{\mu}}$= $9.6\times10^{26}$ s,
Higgs mass $M_{h}$=115 GeV, neutralino mass $M_{\tilde{\chi}}$=1.5$M_{\psi_{\mu}}$
and sneutrino mass $M_{\tilde{\nu}}$=2$M_{\psi_{\mu}}$. In this
parameter set, the $\psi_{\mu}$ decay is dominated by the $Z$ channel
and $W$ channel. Notice that the DM mass here is relatively light,
only $200$ GeV. This is because the decay final states of W and Z
bosons contain plenty of hadrons, especially protons and antiprotons.
However, no excess of antiproton was found by the PAMELA collaboration
\cite{Adriani:2010rc}. In order to avoid the overproduction of anti-proton,
DM with a larger mass is disfavored in this model. Obviously, such
a relatively light DM can not produce electrons and positrons with
energy above $100$ GeV, which falls short of explaining the Fermi-LAT
$e^{\pm}$ excess. To solve this problem, different $e^{\pm}$ backgrounds
have to be considered, as discussed in \cite{Ishiwata:2010am}. The
primary $e^{\pm}$ are supposed to originate from supernova remnant,
which obeys a power law distribution. In \cite{Ishiwata:2010am},
the normalization factor and power index are chosen to be different
from the conventional {}``model 0'' background. This leads to harder
but still plausible backgrounds, which can fit the Fermi-LAT $e^{+}+e^{-}$
spectrum above 100 GeV. Adopting this $e^{\pm}$ backgrounds, the
decaying DM model with R-parity violation could predict $e^{\pm}$
spectra agreeing with the PAMELA and Fermi-LAT observations.

With the same parameter set, the DM induced gamma-ray fluxes from
nearby clusters are predicted to be consistent with the Fermi-LAT
observations, as can be seen from Fig. \ref{fig:RPV}. This is because
the relatively light DM mass $M_{\psi_{\mu}}$= 200 GeV in this scenario
can only produce electrons and positrons with energy below $100$
GeV. That means the ICS process hardly emit gamma-rays above $0.1$
GeV due to kinematics. Then the main contribution to GeV gamma-ray
comes from FSR process in this model. It should be kept in mind that
this interpretation relies on harder $e^{\pm}$ backgrounds. With
this choice of background, one may also decrease the mass of decaying
DM in the framework of GUT to satisfy the gamma-ray constraints from
the clusters. In any case, it predicts the PAMELA positron excess
to be disappear with energy above $100$ GeV, which can definitely
be checked in the near future.

\section{Summary}

In this paper, we have studied the gamma-ray signals from six nearby
clusters of galaxies in decaying DM scenarios. Specifically, we have
concentrated on two decaying DM models which are physically well motivated.
One is decaying DM in the framework of GUT and the other is decaying
DM with R-parity violation. As discussed in previous works, both of
them are able to account for the observational $e^{\pm}$ excesses
and consistent with various measurements of gamma-rays from
inside/outside the Galactic halo. Recently, the Fermi-LAT collaboration
has reported new upper limits on GeV gamma-ray fluxes from clusters
of galaxies. We use these limits to further constrain decaying DM
scenarios.

For GUT decaying DM, we find that the DM induced gamma-ray signals
are comparable in magnitude to the experimental upper limits for all
six clusters. Too much gamma-ray is predicted to come from the M49
and Fornax clusters in the range $1-10$ GeV, which are in disagreement
with Fermi-LAT observations. This conclusion remains unchanged even if we include the
uncertainties of the total DM mass and the starlight in the cluster.
In this energy range most contributions
come from the ICS process, which are model-independent to a large
extent. With the conventional astrophysical background
of electrons and positrons, it seems unlikely for any
decaying DM models to account for the PAMELA and Fermi-LAT excesses
under the constraints of GeV gamma-ray flux upper limits from nearby
clusters. However, it is possible that the Fermi-LAT electrons excess is due
to, for instance, unidentified astrophysical sources instead of DM
decay. In this case, the GUT decaying DM could interpret the PAMELA
positron fraction excess and induce gamma-ray fluxes consistent with
observations from nearby clusters.

Unlike the GUT decaying DM model, the gamma-ray fluxes from nearby
clusters are predicted to be consistent with Fermi-LAT observations
for decaying DM scenario with R-parity violation. This is because
the DM mass here is just around $200$ GeV. With such a relatively
light DM, the ICS process hardly emits gamma-rays above $0.1$ GeV
due to kinematics. The contribution to GeV gamma-ray flux comes
dominantly from the FSR process. It should be kept in mind that in
order to account for the Fermi-LAT $e^{\pm}$ excess in this model,
astrophysical $e^{\pm}$ background harder than conventional one
has to be adopted. As a result, such interpretation would require
the PAMELA positron excess to disappear at energies above $100$
GeV, which could be checked in the near future.
\begin{acknowledgments}
We thank Xiao-Jun Bi for helps on astrophysics and Wei Wu for useful
discussions on PYTHIA package. This work is supported in part by the
National Science Foundation of China (No.10875103, No. 11075139 and
No.10705024) and National Basic Research Program of China (2010CB833000).
M.L and G.Z are also supported in part by the Fundamental Research
Funds for the Central Universities.
\end{acknowledgments}
\bibliographystyle{aipnum4-1}
\bibliography{cluster}

\begin{thebibliography}{49}%
\makeatletter
\providecommand \@ifxundefined [1]{%
 \@ifx{#1\undefined}
}%
\providecommand \@ifnum [1]{%
 \ifnum #1\expandafter \@firstoftwo
 \else \expandafter \@secondoftwo
 \fi
}%
\providecommand \@ifx [1]{%
 \ifx #1\expandafter \@firstoftwo
 \else \expandafter \@secondoftwo
 \fi
}%
\providecommand \natexlab [1]{#1}%
\providecommand \enquote  [1]{``#1''}%
\providecommand \bibnamefont  [1]{#1}%
\providecommand \bibfnamefont [1]{#1}%
\providecommand \citenamefont [1]{#1}%
\providecommand \href@noop [0]{\@secondoftwo}%
\providecommand \href [0]{\begingroup \@sanitize@url \@href}%
\providecommand \@href[1]{\@@startlink{#1}\@@href}%
\providecommand \@@href[1]{\endgroup#1\@@endlink}%
\providecommand \@sanitize@url [0]{\catcode `\\12\catcode `\$12\catcode
  `\&12\catcode `\#12\catcode `\^12\catcode `\_12\catcode `\%12\relax}%
\providecommand \@@startlink[1]{}%
\providecommand \@@endlink[0]{}%
\providecommand \url  [0]{\begingroup\@sanitize@url \@url }%
\providecommand \@url [1]{\endgroup\@href {#1}{\urlprefix }}%
\providecommand \urlprefix  [0]{URL }%
\providecommand \Eprint [0]{\href }%
\@ifxundefined \urlstyle {%
  \providecommand \doi  [0]{\begingroup \@sanitize@url \@doi}%
  \providecommand \@doi [1]{\endgroup \@@startlink {\doibase
  #1}doi:\discretionary {}{}{}#1\@@endlink }%
}{%
  \providecommand \doi  [0]{doi:\discretionary{}{}{}\begingroup
  \urlstyle{rm}\Url }%
}%
\providecommand \doibase [0]{http://dx.doi.org/}%
\providecommand \Doi [0]{\begingroup \@sanitize@url \@Doi }%
\providecommand \@Doi  [1]{\endgroup\@@startlink{\doibase#1}\@@Doi}%
\providecommand \@@Doi [1]{#1\@@endlink}%
\providecommand \selectlanguage [0]{\@gobble}%
\providecommand \bibinfo  [0]{\@secondoftwo}%
\providecommand \bibfield  [0]{\@secondoftwo}%
\providecommand \translation [1]{[#1]}%
\providecommand \BibitemOpen [0]{}%
\providecommand \bibitemStop [0]{}%
\providecommand \bibitemNoStop [0]{.\EOS\space}%
\providecommand \EOS [0]{\spacefactor3000\relax}%
\providecommand \BibitemShut  [1]{\csname bibitem#1\endcsname}%
\bibitem [{\citenamefont {Adriani}\ \emph {et~al.}(2009)\citenamefont {Adriani}
  \emph {et~al.}}]{Adriani:2008zr}%
  \BibitemOpen
  \bibfield  {author} {\bibinfo {author} {\bibfnamefont {O.}~\bibnamefont
  {Adriani}} \emph {et~al.} (\bibinfo {collaboration} {PAMELA}),\ }\Doi
  {10.1038/nature07942} {\bibfield  {journal} {\bibinfo  {journal} {Nature},\
  }\textbf {\bibinfo {volume} {458}},\ \bibinfo {pages} {607} (\bibinfo {year}
  {2009})},\ \Eprint {http://arxiv.org/abs/0810.4995} {arXiv:0810.4995
  [astro-ph]} \BibitemShut {NoStop}%
\bibitem [{\citenamefont {Abdo}\ \emph {et~al.}(2009)\citenamefont {Abdo} \emph
  {et~al.}}]{Abdo:2009zk}%
  \BibitemOpen
  \bibfield  {author} {\bibinfo {author} {\bibfnamefont {A.~A.}\ \bibnamefont
  {Abdo}} \emph {et~al.} (\bibinfo {collaboration} {The Fermi LAT}),\ }\Doi
  {10.1103/PhysRevLett.102.181101} {\bibfield  {journal} {\bibinfo  {journal}
  {Phys. Rev. Lett.},\ }\textbf {\bibinfo {volume} {102}},\ \bibinfo {pages}
  {181101} (\bibinfo {year} {2009})},\ \Eprint {http://arxiv.org/abs/0905.0025}
  {arXiv:0905.0025 [astro-ph.HE]} \BibitemShut {NoStop}%
\bibitem [{\citenamefont {Ackermann}\ \emph
  {et~al.}(2010){\natexlab{a}}\citenamefont {Ackermann} \emph
  {et~al.}}]{Ackermann:2010qj}%
  \BibitemOpen
  \bibfield  {author} {\bibinfo {author} {\bibfnamefont {M.}~\bibnamefont
  {Ackermann}} \emph {et~al.} (\bibinfo {collaboration} {Fermi-LAT}),\
  }\href@noop {} { (\bibinfo {year} {2010}{\natexlab{a}})},\ \Eprint
  {http://arxiv.org/abs/1006.0748} {arXiv:1006.0748 [astro-ph.HE]} \BibitemShut
  {NoStop}%
\bibitem [{\citenamefont {Ackermann}\ \emph
  {et~al.}(2010){\natexlab{b}}\citenamefont {Ackermann} \emph
  {et~al.}}]{Ackermann:2010rg}%
  \BibitemOpen
  \bibfield  {author} {\bibinfo {author} {\bibfnamefont {M.}~\bibnamefont
  {Ackermann}} \emph {et~al.},\ }\Doi {10.1088/1475-7516/2010/05/025}
  {\bibfield  {journal} {\bibinfo  {journal} {JCAP},\ }\textbf {\bibinfo
  {volume} {1005}},\ \bibinfo {pages} {025} (\bibinfo {year}
  {2010}{\natexlab{b}})},\ \Eprint {http://arxiv.org/abs/1002.2239}
  {arXiv:1002.2239 [astro-ph.CO]} \BibitemShut {NoStop}%
\bibitem [{\citenamefont {Yuan}\ \emph {et~al.}(2010)\citenamefont {Yuan},
  \citenamefont {Yin}, \citenamefont {Bi}, \citenamefont {Zhang},\ and\
  \citenamefont {Zhu}}]{Yuan:2010gn}%
  \BibitemOpen
  \bibfield  {author} {\bibinfo {author} {\bibfnamefont {Q.}~\bibnamefont
  {Yuan}}, \bibinfo {author} {\bibfnamefont {P.-F.}\ \bibnamefont {Yin}},
  \bibinfo {author} {\bibfnamefont {X.-J.}\ \bibnamefont {Bi}}, \bibinfo
  {author} {\bibfnamefont {X.-M.}\ \bibnamefont {Zhang}}, \ and\ \bibinfo
  {author} {\bibfnamefont {S.-H.}\ \bibnamefont {Zhu}},\ }\Doi
  {10.1103/PhysRevD.82.023506} {\bibfield  {journal} {\bibinfo  {journal}
  {Phys. Rev.},\ }\textbf {\bibinfo {volume} {D82}},\ \bibinfo {pages} {023506}
  (\bibinfo {year} {2010})},\ \Eprint {http://arxiv.org/abs/1002.0197}
  {arXiv:1002.0197 [astro-ph.HE]} \BibitemShut {NoStop}%
\bibitem [{\citenamefont {Dugger}\ \emph {et~al.}(2010)\citenamefont {Dugger},
  \citenamefont {Jeltema},\ and\ \citenamefont {Profumo}}]{Dugger:2010ys}%
  \BibitemOpen
  \bibfield  {author} {\bibinfo {author} {\bibfnamefont {L.}~\bibnamefont
  {Dugger}}, \bibinfo {author} {\bibfnamefont {T.~E.}\ \bibnamefont {Jeltema}},
  \ and\ \bibinfo {author} {\bibfnamefont {S.}~\bibnamefont {Profumo}},\ }\Doi
  {10.1088/1475-7516/2010/12/015} {\bibfield  {journal} {\bibinfo  {journal}
  {JCAP},\ }\textbf {\bibinfo {volume} {1012}},\ \bibinfo {pages} {015}
  (\bibinfo {year} {2010})},\ \Eprint {http://arxiv.org/abs/1009.5988}
  {arXiv:1009.5988 [astro-ph.HE]} \BibitemShut {NoStop}%
\bibitem [{\citenamefont {Bergstrom}\ \emph {et~al.}(2008)\citenamefont
  {Bergstrom}, \citenamefont {Bringmann},\ and\ \citenamefont
  {Edsjo}}]{Bergstrom:2008gr}%
  \BibitemOpen
  \bibfield  {author} {\bibinfo {author} {\bibfnamefont {L.}~\bibnamefont
  {Bergstrom}}, \bibinfo {author} {\bibfnamefont {T.}~\bibnamefont
  {Bringmann}}, \ and\ \bibinfo {author} {\bibfnamefont {J.}~\bibnamefont
  {Edsjo}},\ }\Doi {10.1103/PhysRevD.78.103520} {\bibfield  {journal} {\bibinfo
   {journal} {Phys. Rev.},\ }\textbf {\bibinfo {volume} {D78}},\ \bibinfo
  {pages} {103520} (\bibinfo {year} {2008})},\ \Eprint
  {http://arxiv.org/abs/0808.3725} {arXiv:0808.3725 [astro-ph]} \BibitemShut
  {NoStop}%
\bibitem [{\citenamefont {Cholis}\ \emph {et~al.}(2009)\citenamefont {Cholis},
  \citenamefont {Goodenough}, \citenamefont {Hooper}, \citenamefont {Simet},\
  and\ \citenamefont {Weiner}}]{Cholis:2008hb}%
  \BibitemOpen
  \bibfield  {author} {\bibinfo {author} {\bibfnamefont {I.}~\bibnamefont
  {Cholis}}, \bibinfo {author} {\bibfnamefont {L.}~\bibnamefont {Goodenough}},
  \bibinfo {author} {\bibfnamefont {D.}~\bibnamefont {Hooper}}, \bibinfo
  {author} {\bibfnamefont {M.}~\bibnamefont {Simet}}, \ and\ \bibinfo {author}
  {\bibfnamefont {N.}~\bibnamefont {Weiner}},\ }\Doi
  {10.1103/PhysRevD.80.123511} {\bibfield  {journal} {\bibinfo  {journal}
  {Phys. Rev.},\ }\textbf {\bibinfo {volume} {D80}},\ \bibinfo {pages} {123511}
  (\bibinfo {year} {2009})},\ \Eprint {http://arxiv.org/abs/0809.1683}
  {arXiv:0809.1683 [hep-ph]} \BibitemShut {NoStop}%
\bibitem [{\citenamefont {Cirelli}\ \emph {et~al.}(2009)\citenamefont
  {Cirelli}, \citenamefont {Kadastik}, \citenamefont {Raidal},\ and\
  \citenamefont {Strumia}}]{Cirelli:2008pk}%
  \BibitemOpen
  \bibfield  {author} {\bibinfo {author} {\bibfnamefont {M.}~\bibnamefont
  {Cirelli}}, \bibinfo {author} {\bibfnamefont {M.}~\bibnamefont {Kadastik}},
  \bibinfo {author} {\bibfnamefont {M.}~\bibnamefont {Raidal}}, \ and\ \bibinfo
  {author} {\bibfnamefont {A.}~\bibnamefont {Strumia}},\ }\Doi
  {10.1016/j.nuclphysb.2008.11.031} {\bibfield  {journal} {\bibinfo  {journal}
  {Nucl. Phys.},\ }\textbf {\bibinfo {volume} {B813}},\ \bibinfo {pages} {1}
  (\bibinfo {year} {2009})},\ \Eprint {http://arxiv.org/abs/0809.2409}
  {arXiv:0809.2409 [hep-ph]} \BibitemShut {NoStop}%
\bibitem [{\citenamefont {Nardi}\ \emph {et~al.}(2009)\citenamefont {Nardi},
  \citenamefont {Sannino},\ and\ \citenamefont {Strumia}}]{Nardi:2008ix}%
  \BibitemOpen
  \bibfield  {author} {\bibinfo {author} {\bibfnamefont {E.}~\bibnamefont
  {Nardi}}, \bibinfo {author} {\bibfnamefont {F.}~\bibnamefont {Sannino}}, \
  and\ \bibinfo {author} {\bibfnamefont {A.}~\bibnamefont {Strumia}},\ }\Doi
  {10.1088/1475-7516/2009/01/043} {\bibfield  {journal} {\bibinfo  {journal}
  {JCAP},\ }\textbf {\bibinfo {volume} {0901}},\ \bibinfo {pages} {043}
  (\bibinfo {year} {2009})},\ \Eprint {http://arxiv.org/abs/0811.4153}
  {arXiv:0811.4153 [hep-ph]} \BibitemShut {NoStop}%
\bibitem [{\citenamefont {Bertone}\ \emph {et~al.}(2009)\citenamefont
  {Bertone}, \citenamefont {Cirelli}, \citenamefont {Strumia},\ and\
  \citenamefont {Taoso}}]{Bertone:2008xr}%
  \BibitemOpen
  \bibfield  {author} {\bibinfo {author} {\bibfnamefont {G.}~\bibnamefont
  {Bertone}}, \bibinfo {author} {\bibfnamefont {M.}~\bibnamefont {Cirelli}},
  \bibinfo {author} {\bibfnamefont {A.}~\bibnamefont {Strumia}}, \ and\
  \bibinfo {author} {\bibfnamefont {M.}~\bibnamefont {Taoso}},\ }\Doi
  {10.1088/1475-7516/2009/03/009} {\bibfield  {journal} {\bibinfo  {journal}
  {JCAP},\ }\textbf {\bibinfo {volume} {0903}},\ \bibinfo {pages} {009}
  (\bibinfo {year} {2009})},\ \Eprint {http://arxiv.org/abs/0811.3744}
  {arXiv:0811.3744 [astro-ph]} \BibitemShut {NoStop}%
\bibitem [{\citenamefont {Cirelli}\ \emph {et~al.}(2010)\citenamefont
  {Cirelli}, \citenamefont {Panci},\ and\ \citenamefont
  {Serpico}}]{Cirelli:2009dv}%
  \BibitemOpen
  \bibfield  {author} {\bibinfo {author} {\bibfnamefont {M.}~\bibnamefont
  {Cirelli}}, \bibinfo {author} {\bibfnamefont {P.}~\bibnamefont {Panci}}, \
  and\ \bibinfo {author} {\bibfnamefont {P.~D.}\ \bibnamefont {Serpico}},\
  }\Doi {10.1016/j.nuclphysb.2010.07.010} {\bibfield  {journal} {\bibinfo
  {journal} {Nucl. Phys.},\ }\textbf {\bibinfo {volume} {B840}},\ \bibinfo
  {pages} {284} (\bibinfo {year} {2010})},\ \Eprint
  {http://arxiv.org/abs/0912.0663} {arXiv:0912.0663 [astro-ph.CO]} \BibitemShut
  {NoStop}%
\bibitem [{\citenamefont {Papucci}\ and\ \citenamefont
  {Strumia}(2010)}]{Papucci:2009gd}%
  \BibitemOpen
  \bibfield  {author} {\bibinfo {author} {\bibfnamefont {M.}~\bibnamefont
  {Papucci}}\ and\ \bibinfo {author} {\bibfnamefont {A.}~\bibnamefont
  {Strumia}},\ }\Doi {10.1088/1475-7516/2010/03/014} {\bibfield  {journal}
  {\bibinfo  {journal} {JCAP},\ }\textbf {\bibinfo {volume} {1003}},\ \bibinfo
  {pages} {014} (\bibinfo {year} {2010})},\ \Eprint
  {http://arxiv.org/abs/0912.0742} {arXiv:0912.0742 [hep-ph]} \BibitemShut
  {NoStop}%
\bibitem [{\citenamefont {Liu}\ \emph {et~al.}(2011)\citenamefont {Liu},
  \citenamefont {Wu},\ and\ \citenamefont {Zhou}}]{Liu:2011aa}%
  \BibitemOpen
  \bibfield  {author} {\bibinfo {author} {\bibfnamefont {Z.-P.}\ \bibnamefont
  {Liu}}, \bibinfo {author} {\bibfnamefont {Y.-L.}\ \bibnamefont {Wu}}, \ and\
  \bibinfo {author} {\bibfnamefont {Y.-F.}\ \bibnamefont {Zhou}},\ }\href@noop
  {} { (\bibinfo {year} {2011})},\ \Eprint {http://arxiv.org/abs/1101.4148}
  {arXiv:1101.4148 [hep-ph]} \BibitemShut {NoStop}%
\bibitem [{\citenamefont {Carone}\ \emph {et~al.}(2010)\citenamefont {Carone},
  \citenamefont {Erlich},\ and\ \citenamefont {Primulando}}]{Carone:2010ha}%
  \BibitemOpen
  \bibfield  {author} {\bibinfo {author} {\bibfnamefont {C.~D.}\ \bibnamefont
  {Carone}}, \bibinfo {author} {\bibfnamefont {J.}~\bibnamefont {Erlich}}, \
  and\ \bibinfo {author} {\bibfnamefont {R.}~\bibnamefont {Primulando}},\ }\Doi
  {10.1103/PhysRevD.82.055028} {\bibfield  {journal} {\bibinfo  {journal}
  {Phys. Rev.},\ }\textbf {\bibinfo {volume} {D82}},\ \bibinfo {pages} {055028}
  (\bibinfo {year} {2010})},\ \Eprint {http://arxiv.org/abs/1008.0642}
  {arXiv:1008.0642 [hep-ph]} \BibitemShut {NoStop}%
\bibitem [{\citenamefont {Cheng}\ \emph {et~al.}(2010)\citenamefont {Cheng},
  \citenamefont {Huang}, \citenamefont {Low},\ and\ \citenamefont
  {Menon}}]{Cheng:2010mw}%
  \BibitemOpen
  \bibfield  {author} {\bibinfo {author} {\bibfnamefont {H.-C.}\ \bibnamefont
  {Cheng}}, \bibinfo {author} {\bibfnamefont {W.-C.}\ \bibnamefont {Huang}},
  \bibinfo {author} {\bibfnamefont {I.}~\bibnamefont {Low}}, \ and\ \bibinfo
  {author} {\bibfnamefont {A.}~\bibnamefont {Menon}},\ }\href@noop {} {
  (\bibinfo {year} {2010})},\ \Eprint {http://arxiv.org/abs/1012.5300}
  {arXiv:1012.5300 [hep-ph]} \BibitemShut {NoStop}%
\bibitem [{\citenamefont {Jeltema}\ \emph {et~al.}(2009)\citenamefont
  {Jeltema}, \citenamefont {Kehayias},\ and\ \citenamefont
  {Profumo}}]{Jeltema:2008vu}%
  \BibitemOpen
  \bibfield  {author} {\bibinfo {author} {\bibfnamefont {T.~E.}\ \bibnamefont
  {Jeltema}}, \bibinfo {author} {\bibfnamefont {J.}~\bibnamefont {Kehayias}}, \
  and\ \bibinfo {author} {\bibfnamefont {S.}~\bibnamefont {Profumo}},\ }\Doi
  {10.1103/PhysRevD.80.023005} {\bibfield  {journal} {\bibinfo  {journal}
  {Phys. Rev.},\ }\textbf {\bibinfo {volume} {D80}},\ \bibinfo {pages} {023005}
  (\bibinfo {year} {2009})},\ \Eprint {http://arxiv.org/abs/0812.0597}
  {arXiv:0812.0597 [astro-ph]} \BibitemShut {NoStop}%
\bibitem [{\citenamefont {Pinzke}\ \emph {et~al.}(2009)\citenamefont {Pinzke},
  \citenamefont {Pfrommer},\ and\ \citenamefont {Bergstrom}}]{Pinzke:2009cp}%
  \BibitemOpen
  \bibfield  {author} {\bibinfo {author} {\bibfnamefont {A.}~\bibnamefont
  {Pinzke}}, \bibinfo {author} {\bibfnamefont {C.}~\bibnamefont {Pfrommer}}, \
  and\ \bibinfo {author} {\bibfnamefont {L.}~\bibnamefont {Bergstrom}},\ }\Doi
  {10.1103/PhysRevLett.103.181302} {\bibfield  {journal} {\bibinfo  {journal}
  {Phys. Rev. Lett.},\ }\textbf {\bibinfo {volume} {103}},\ \bibinfo {pages}
  {181302} (\bibinfo {year} {2009})},\ \Eprint {http://arxiv.org/abs/0905.1948}
  {arXiv:0905.1948 [astro-ph.HE]} \BibitemShut {NoStop}%
\bibitem [{\citenamefont {Cuesta}\ \emph {et~al.}(2010)\citenamefont {Cuesta}
  \emph {et~al.}}]{Cuesta:2010ex}%
  \BibitemOpen
  \bibfield  {author} {\bibinfo {author} {\bibfnamefont {A.~J.}\ \bibnamefont
  {Cuesta}} \emph {et~al.},\ }\href@noop {} { (\bibinfo {year} {2010})},\
  \Eprint {http://arxiv.org/abs/1007.3469} {arXiv:1007.3469 [astro-ph.HE]}
  \BibitemShut {NoStop}%
\bibitem [{\citenamefont {Reiprich}\ and\ \citenamefont
  {Boehringer}(2002)}]{Reiprich:2001zv}%
  \BibitemOpen
  \bibfield  {author} {\bibinfo {author} {\bibfnamefont {T.~H.}\ \bibnamefont
  {Reiprich}}\ and\ \bibinfo {author} {\bibfnamefont {H.}~\bibnamefont
  {Boehringer}},\ }\Doi {10.1086/338753} {\bibfield  {journal} {\bibinfo
  {journal} {Astrophys. J.},\ }\textbf {\bibinfo {volume} {567}},\ \bibinfo
  {pages} {716} (\bibinfo {year} {2002})},\ \Eprint
  {http://arxiv.org/abs/astro-ph/0111285} {arXiv:astro-ph/0111285} \BibitemShut
  {NoStop}%
\bibitem [{\citenamefont {Sjostrand}\ \emph {et~al.}(2006)\citenamefont
  {Sjostrand}, \citenamefont {Mrenna},\ and\ \citenamefont
  {Skands}}]{Sjostrand:2006za}%
  \BibitemOpen
  \bibfield  {author} {\bibinfo {author} {\bibfnamefont {T.}~\bibnamefont
  {Sjostrand}}, \bibinfo {author} {\bibfnamefont {S.}~\bibnamefont {Mrenna}}, \
  and\ \bibinfo {author} {\bibfnamefont {P.}~\bibnamefont {Skands}},\
  }\href@noop {} {\bibfield  {journal} {\bibinfo  {journal} {JHEP},\ }\textbf
  {\bibinfo {volume} {05}},\ \bibinfo {pages} {026} (\bibinfo {year} {2006})},\
  \Eprint {http://arxiv.org/abs/hep-ph/0603175} {arXiv:hep-ph/0603175}
  \BibitemShut {NoStop}%
\bibitem [{\citenamefont {Blumenthal}\ and\ \citenamefont
  {Gould}(1970)}]{Blumenthal:1970gc}%
  \BibitemOpen
  \bibfield  {author} {\bibinfo {author} {\bibfnamefont {G.~R.}\ \bibnamefont
  {Blumenthal}}\ and\ \bibinfo {author} {\bibfnamefont {R.~J.}\ \bibnamefont
  {Gould}},\ }\Doi {10.1103/RevModPhys.42.237} {\bibfield  {journal} {\bibinfo
  {journal} {Rev. Mod. Phys.},\ }\textbf {\bibinfo {volume} {42}},\ \bibinfo
  {pages} {237} (\bibinfo {year} {1970})}\BibitemShut {NoStop}%
\bibitem [{\citenamefont {Cirelli}\ and\ \citenamefont
  {Panci}(2009)}]{Cirelli:2009vg}%
  \BibitemOpen
  \bibfield  {author} {\bibinfo {author} {\bibfnamefont {M.}~\bibnamefont
  {Cirelli}}\ and\ \bibinfo {author} {\bibfnamefont {P.}~\bibnamefont
  {Panci}},\ }\Doi {10.1016/j.nuclphysb.2009.06.034} {\bibfield  {journal}
  {\bibinfo  {journal} {Nucl. Phys.},\ }\textbf {\bibinfo {volume} {B821}},\
  \bibinfo {pages} {399} (\bibinfo {year} {2009})},\ \Eprint
  {http://arxiv.org/abs/0904.3830} {arXiv:0904.3830 [astro-ph.CO]} \BibitemShut
  {NoStop}%
\bibitem [{\citenamefont {Ibarra}\ \emph
  {et~al.}(2010){\natexlab{a}}\citenamefont {Ibarra}, \citenamefont {Tran},\
  and\ \citenamefont {Weniger}}]{Ibarra:2009dr}%
  \BibitemOpen
  \bibfield  {author} {\bibinfo {author} {\bibfnamefont {A.}~\bibnamefont
  {Ibarra}}, \bibinfo {author} {\bibfnamefont {D.}~\bibnamefont {Tran}}, \ and\
  \bibinfo {author} {\bibfnamefont {C.}~\bibnamefont {Weniger}},\ }\Doi
  {10.1088/1475-7516/2010/01/009} {\bibfield  {journal} {\bibinfo  {journal}
  {JCAP},\ }\textbf {\bibinfo {volume} {1001}},\ \bibinfo {pages} {009}
  (\bibinfo {year} {2010}{\natexlab{a}})},\ \Eprint
  {http://arxiv.org/abs/0906.1571} {arXiv:0906.1571 [hep-ph]} \BibitemShut
  {NoStop}%
\bibitem [{\citenamefont {Ibarra}\ \emph
  {et~al.}(2010){\natexlab{b}}\citenamefont {Ibarra}, \citenamefont {Tran},\
  and\ \citenamefont {Weniger}}]{Ibarra:2009nw}%
  \BibitemOpen
  \bibfield  {author} {\bibinfo {author} {\bibfnamefont {A.}~\bibnamefont
  {Ibarra}}, \bibinfo {author} {\bibfnamefont {D.}~\bibnamefont {Tran}}, \ and\
  \bibinfo {author} {\bibfnamefont {C.}~\bibnamefont {Weniger}},\ }\Doi
  {10.1103/PhysRevD.81.023529} {\bibfield  {journal} {\bibinfo  {journal}
  {Phys. Rev.},\ }\textbf {\bibinfo {volume} {D81}},\ \bibinfo {pages} {023529}
  (\bibinfo {year} {2010}{\natexlab{b}})},\ \Eprint
  {http://arxiv.org/abs/0909.3514} {arXiv:0909.3514 [hep-ph]} \BibitemShut
  {NoStop}%
\bibitem [{\citenamefont {Ishiwata}\ \emph
  {et~al.}(2009){\natexlab{a}}\citenamefont {Ishiwata}, \citenamefont
  {Matsumoto},\ and\ \citenamefont {Moroi}}]{Ishiwata:2009dk}%
  \BibitemOpen
  \bibfield  {author} {\bibinfo {author} {\bibfnamefont {K.}~\bibnamefont
  {Ishiwata}}, \bibinfo {author} {\bibfnamefont {S.}~\bibnamefont {Matsumoto}},
  \ and\ \bibinfo {author} {\bibfnamefont {T.}~\bibnamefont {Moroi}},\ }\Doi
  {10.1016/j.physletb.2009.07.004} {\bibfield  {journal} {\bibinfo  {journal}
  {Phys. Lett.},\ }\textbf {\bibinfo {volume} {B679}},\ \bibinfo {pages} {1}
  (\bibinfo {year} {2009}{\natexlab{a}})},\ \Eprint
  {http://arxiv.org/abs/0905.4593} {arXiv:0905.4593 [astro-ph.CO]} \BibitemShut
  {NoStop}%
\bibitem [{\citenamefont {Luo}\ \emph {et~al.}(2010)\citenamefont {Luo},
  \citenamefont {Wang}, \citenamefont {Wu},\ and\ \citenamefont
  {Zhu}}]{Luo:2009xd}%
  \BibitemOpen
  \bibfield  {author} {\bibinfo {author} {\bibfnamefont {M.}~\bibnamefont
  {Luo}}, \bibinfo {author} {\bibfnamefont {L.}~\bibnamefont {Wang}}, \bibinfo
  {author} {\bibfnamefont {W.}~\bibnamefont {Wu}}, \ and\ \bibinfo {author}
  {\bibfnamefont {G.}~\bibnamefont {Zhu}},\ }\Doi
  {10.1016/j.physletb.2010.04.014} {\bibfield  {journal} {\bibinfo  {journal}
  {Phys. Lett.},\ }\textbf {\bibinfo {volume} {B688}},\ \bibinfo {pages} {216}
  (\bibinfo {year} {2010})},\ \Eprint {http://arxiv.org/abs/0911.3235}
  {arXiv:0911.3235 [hep-ph]} \BibitemShut {NoStop}%
\bibitem [{\citenamefont {Zhang}\ \emph {et~al.}(2010)\citenamefont {Zhang},
  \citenamefont {Yuan},\ and\ \citenamefont {Bi}}]{Zhang:2009kp}%
  \BibitemOpen
  \bibfield  {author} {\bibinfo {author} {\bibfnamefont {J.}~\bibnamefont
  {Zhang}}, \bibinfo {author} {\bibfnamefont {Q.}~\bibnamefont {Yuan}}, \ and\
  \bibinfo {author} {\bibfnamefont {X.-J.}\ \bibnamefont {Bi}},\ }\Doi
  {10.1088/0004-637X/720/1/9} {\bibfield  {journal} {\bibinfo  {journal}
  {Astrophys. J.},\ }\textbf {\bibinfo {volume} {720}},\ \bibinfo {pages} {9}
  (\bibinfo {year} {2010})},\ \Eprint {http://arxiv.org/abs/0908.1236}
  {arXiv:0908.1236 [astro-ph.HE]} \BibitemShut {NoStop}%
\bibitem [{\citenamefont {Chen}\ \emph
  {et~al.}(2009){\natexlab{a}}\citenamefont {Chen}, \citenamefont {Takahashi},\
  and\ \citenamefont {Yanagida}}]{Chen:2008md}%
  \BibitemOpen
  \bibfield  {author} {\bibinfo {author} {\bibfnamefont {C.-R.}\ \bibnamefont
  {Chen}}, \bibinfo {author} {\bibfnamefont {F.}~\bibnamefont {Takahashi}}, \
  and\ \bibinfo {author} {\bibfnamefont {T.~T.}\ \bibnamefont {Yanagida}},\
  }\Doi {10.1016/j.physletb.2009.02.046} {\bibfield  {journal} {\bibinfo
  {journal} {Phys. Lett.},\ }\textbf {\bibinfo {volume} {B673}},\ \bibinfo
  {pages} {255} (\bibinfo {year} {2009}{\natexlab{a}})},\ \Eprint
  {http://arxiv.org/abs/0811.0477} {arXiv:0811.0477 [hep-ph]} \BibitemShut
  {NoStop}%
\bibitem [{\citenamefont {Arvanitaki}\ \emph
  {et~al.}(2009){\natexlab{a}}\citenamefont {Arvanitaki} \emph
  {et~al.}}]{Arvanitaki:2008hq}%
  \BibitemOpen
  \bibfield  {author} {\bibinfo {author} {\bibfnamefont {A.}~\bibnamefont
  {Arvanitaki}} \emph {et~al.},\ }\Doi {10.1103/PhysRevD.79.105022} {\bibfield
  {journal} {\bibinfo  {journal} {Phys. Rev.},\ }\textbf {\bibinfo {volume}
  {D79}},\ \bibinfo {pages} {105022} (\bibinfo {year} {2009}{\natexlab{a}})},\
  \Eprint {http://arxiv.org/abs/0812.2075} {arXiv:0812.2075 [hep-ph]}
  \BibitemShut {NoStop}%
\bibitem [{\citenamefont {Arvanitaki}\ \emph
  {et~al.}(2009){\natexlab{b}}\citenamefont {Arvanitaki} \emph
  {et~al.}}]{Arvanitaki:2009yb}%
  \BibitemOpen
  \bibfield  {author} {\bibinfo {author} {\bibfnamefont {A.}~\bibnamefont
  {Arvanitaki}} \emph {et~al.},\ }\Doi {10.1103/PhysRevD.80.055011} {\bibfield
  {journal} {\bibinfo  {journal} {Phys. Rev.},\ }\textbf {\bibinfo {volume}
  {D80}},\ \bibinfo {pages} {055011} (\bibinfo {year} {2009}{\natexlab{b}})},\
  \Eprint {http://arxiv.org/abs/0904.2789} {arXiv:0904.2789 [hep-ph]}
  \BibitemShut {NoStop}%
\bibitem [{\citenamefont {Ruderman}\ and\ \citenamefont
  {Volansky}(2009)}]{Ruderman:2009ta}%
  \BibitemOpen
  \bibfield  {author} {\bibinfo {author} {\bibfnamefont {J.~T.}\ \bibnamefont
  {Ruderman}}\ and\ \bibinfo {author} {\bibfnamefont {T.}~\bibnamefont
  {Volansky}},\ }\href@noop {} { (\bibinfo {year} {2009})},\ \Eprint
  {http://arxiv.org/abs/0907.4373} {arXiv:0907.4373 [hep-ph]} \BibitemShut
  {NoStop}%
\bibitem [{\citenamefont {Ruderman}\ and\ \citenamefont
  {Volansky}(2010)}]{Ruderman:2009tj}%
  \BibitemOpen
  \bibfield  {author} {\bibinfo {author} {\bibfnamefont {J.~T.}\ \bibnamefont
  {Ruderman}}\ and\ \bibinfo {author} {\bibfnamefont {T.}~\bibnamefont
  {Volansky}},\ }\Doi {10.1007/JHEP02(2010)024} {\bibfield  {journal} {\bibinfo
   {journal} {JHEP},\ }\textbf {\bibinfo {volume} {02}},\ \bibinfo {pages}
  {024} (\bibinfo {year} {2010})},\ \Eprint {http://arxiv.org/abs/0908.1570}
  {arXiv:0908.1570 [hep-ph]} \BibitemShut {NoStop}%
\bibitem [{\citenamefont {Kadastik}\ \emph {et~al.}(2009)\citenamefont
  {Kadastik}, \citenamefont {Kannike},\ and\ \citenamefont
  {Raidal}}]{Kadastik:2009cu}%
  \BibitemOpen
  \bibfield  {author} {\bibinfo {author} {\bibfnamefont {M.}~\bibnamefont
  {Kadastik}}, \bibinfo {author} {\bibfnamefont {K.}~\bibnamefont {Kannike}}, \
  and\ \bibinfo {author} {\bibfnamefont {M.}~\bibnamefont {Raidal}},\ }\Doi
  {10.1103/PhysRevD.80.085020} {\bibfield  {journal} {\bibinfo  {journal}
  {Phys. Rev.},\ }\textbf {\bibinfo {volume} {D80}},\ \bibinfo {pages} {085020}
  (\bibinfo {year} {2009})},\ \Eprint {http://arxiv.org/abs/0907.1894}
  {arXiv:0907.1894 [hep-ph]} \BibitemShut {NoStop}%
\bibitem [{\citenamefont {Kyae}(2010)}]{Kyae:2009gm}%
  \BibitemOpen
  \bibfield  {author} {\bibinfo {author} {\bibfnamefont {B.}~\bibnamefont
  {Kyae}},\ }\Doi {10.1016/j.physletb.2009.12.068} {\bibfield  {journal}
  {\bibinfo  {journal} {Phys. Lett.},\ }\textbf {\bibinfo {volume} {B685}},\
  \bibinfo {pages} {19} (\bibinfo {year} {2010})},\ \Eprint
  {http://arxiv.org/abs/0909.3139} {arXiv:0909.3139 [hep-ph]} \BibitemShut
  {NoStop}%
\bibitem [{\citenamefont {Haba}\ \emph {et~al.}(2011)\citenamefont {Haba},
  \citenamefont {Kajiyama}, \citenamefont {Matsumoto}, \citenamefont {Okada},\
  and\ \citenamefont {Yoshioka}}]{Haba:2010ag}%
  \BibitemOpen
  \bibfield  {author} {\bibinfo {author} {\bibfnamefont {N.}~\bibnamefont
  {Haba}}, \bibinfo {author} {\bibfnamefont {Y.}~\bibnamefont {Kajiyama}},
  \bibinfo {author} {\bibfnamefont {S.}~\bibnamefont {Matsumoto}}, \bibinfo
  {author} {\bibfnamefont {H.}~\bibnamefont {Okada}}, \ and\ \bibinfo {author}
  {\bibfnamefont {K.}~\bibnamefont {Yoshioka}},\ }\Doi
  {10.1016/j.physletb.2010.11.063} {\bibfield  {journal} {\bibinfo  {journal}
  {Phys. Lett.},\ }\textbf {\bibinfo {volume} {B695}},\ \bibinfo {pages} {476}
  (\bibinfo {year} {2011})},\ \Eprint {http://arxiv.org/abs/1008.4777}
  {arXiv:1008.4777 [hep-ph]} \BibitemShut {NoStop}%
\bibitem [{\citenamefont {Gao}\ \emph {et~al.}(2010)\citenamefont {Gao},
  \citenamefont {Kang},\ and\ \citenamefont {Li}}]{Gao:2010pg}%
  \BibitemOpen
  \bibfield  {author} {\bibinfo {author} {\bibfnamefont {X.}~\bibnamefont
  {Gao}}, \bibinfo {author} {\bibfnamefont {Z.}~\bibnamefont {Kang}}, \ and\
  \bibinfo {author} {\bibfnamefont {T.}~\bibnamefont {Li}},\ }\Doi
  {10.1140/epjc/s10052-010-1418-z} {\bibfield  {journal} {\bibinfo  {journal}
  {Eur. Phys. J.},\ }\textbf {\bibinfo {volume} {C69}},\ \bibinfo {pages} {467}
  (\bibinfo {year} {2010})},\ \Eprint {http://arxiv.org/abs/1001.3278}
  {arXiv:1001.3278 [hep-ph]} \BibitemShut {NoStop}%
\bibitem [{\citenamefont {Buchmuller}\ \emph {et~al.}(2007)\citenamefont
  {Buchmuller}, \citenamefont {Covi}, \citenamefont {Hamaguchi}, \citenamefont
  {Ibarra},\ and\ \citenamefont {Yanagida}}]{Buchmuller:2007ui}%
  \BibitemOpen
  \bibfield  {author} {\bibinfo {author} {\bibfnamefont {W.}~\bibnamefont
  {Buchmuller}}, \bibinfo {author} {\bibfnamefont {L.}~\bibnamefont {Covi}},
  \bibinfo {author} {\bibfnamefont {K.}~\bibnamefont {Hamaguchi}}, \bibinfo
  {author} {\bibfnamefont {A.}~\bibnamefont {Ibarra}}, \ and\ \bibinfo {author}
  {\bibfnamefont {T.}~\bibnamefont {Yanagida}},\ }\Doi
  {10.1088/1126-6708/2007/03/037} {\bibfield  {journal} {\bibinfo  {journal}
  {JHEP},\ }\textbf {\bibinfo {volume} {03}},\ \bibinfo {pages} {037} (\bibinfo
  {year} {2007})},\ \Eprint {http://arxiv.org/abs/hep-ph/0702184}
  {arXiv:hep-ph/0702184} \BibitemShut {NoStop}%
\bibitem [{\citenamefont {Ishiwata}\ \emph {et~al.}(2008)\citenamefont
  {Ishiwata}, \citenamefont {Matsumoto},\ and\ \citenamefont
  {Moroi}}]{Ishiwata:2008cu}%
  \BibitemOpen
  \bibfield  {author} {\bibinfo {author} {\bibfnamefont {K.}~\bibnamefont
  {Ishiwata}}, \bibinfo {author} {\bibfnamefont {S.}~\bibnamefont {Matsumoto}},
  \ and\ \bibinfo {author} {\bibfnamefont {T.}~\bibnamefont {Moroi}},\ }\Doi
  {10.1103/PhysRevD.78.063505} {\bibfield  {journal} {\bibinfo  {journal}
  {Phys. Rev.},\ }\textbf {\bibinfo {volume} {D78}},\ \bibinfo {pages} {063505}
  (\bibinfo {year} {2008})},\ \Eprint {http://arxiv.org/abs/0805.1133}
  {arXiv:0805.1133 [hep-ph]} \BibitemShut {NoStop}%
\bibitem [{\citenamefont {Ishiwata}\ \emph {et~al.}(2010)\citenamefont
  {Ishiwata}, \citenamefont {Matsumoto},\ and\ \citenamefont
  {Moroi}}]{Ishiwata:2010am}%
  \BibitemOpen
  \bibfield  {author} {\bibinfo {author} {\bibfnamefont {K.}~\bibnamefont
  {Ishiwata}}, \bibinfo {author} {\bibfnamefont {S.}~\bibnamefont {Matsumoto}},
  \ and\ \bibinfo {author} {\bibfnamefont {T.}~\bibnamefont {Moroi}},\ }\Doi
  {10.1007/JHEP12(2010)006} {\bibfield  {journal} {\bibinfo  {journal} {JHEP},\
  }\textbf {\bibinfo {volume} {12}},\ \bibinfo {pages} {006} (\bibinfo {year}
  {2010})},\ \Eprint {http://arxiv.org/abs/1008.3636} {arXiv:1008.3636
  [hep-ph]} \BibitemShut {NoStop}%
\bibitem [{\citenamefont {Ishiwata}\ \emph
  {et~al.}(2009){\natexlab{b}}\citenamefont {Ishiwata}, \citenamefont
  {Matsumoto},\ and\ \citenamefont {Moroi}}]{Ishiwata:2009vx}%
  \BibitemOpen
  \bibfield  {author} {\bibinfo {author} {\bibfnamefont {K.}~\bibnamefont
  {Ishiwata}}, \bibinfo {author} {\bibfnamefont {S.}~\bibnamefont {Matsumoto}},
  \ and\ \bibinfo {author} {\bibfnamefont {T.}~\bibnamefont {Moroi}},\ }\Doi
  {10.1088/1126-6708/2009/05/110} {\bibfield  {journal} {\bibinfo  {journal}
  {JHEP},\ }\textbf {\bibinfo {volume} {05}},\ \bibinfo {pages} {110} (\bibinfo
  {year} {2009}{\natexlab{b}})},\ \Eprint {http://arxiv.org/abs/0903.0242}
  {arXiv:0903.0242 [hep-ph]} \BibitemShut {NoStop}%
\bibitem [{\citenamefont {Shirai}\ \emph {et~al.}(2009)\citenamefont {Shirai},
  \citenamefont {Takahashi},\ and\ \citenamefont {Yanagida}}]{Shirai:2009fq}%
  \BibitemOpen
  \bibfield  {author} {\bibinfo {author} {\bibfnamefont {S.}~\bibnamefont
  {Shirai}}, \bibinfo {author} {\bibfnamefont {F.}~\bibnamefont {Takahashi}}, \
  and\ \bibinfo {author} {\bibfnamefont {T.~T.}\ \bibnamefont {Yanagida}},\
  }\Doi {10.1016/j.physletb.2009.09.049} {\bibfield  {journal} {\bibinfo
  {journal} {Phys. Lett.},\ }\textbf {\bibinfo {volume} {B680}},\ \bibinfo
  {pages} {485} (\bibinfo {year} {2009})},\ \Eprint
  {http://arxiv.org/abs/0905.0388} {arXiv:0905.0388 [hep-ph]} \BibitemShut
  {NoStop}%
\bibitem [{\citenamefont {Chen}\ \emph
  {et~al.}(2009){\natexlab{b}}\citenamefont {Chen}, \citenamefont {Mohapatra},
  \citenamefont {Nussinov},\ and\ \citenamefont {Zhang}}]{Chen:2009ew}%
  \BibitemOpen
  \bibfield  {author} {\bibinfo {author} {\bibfnamefont {S.-L.}\ \bibnamefont
  {Chen}}, \bibinfo {author} {\bibfnamefont {R.~N.}\ \bibnamefont {Mohapatra}},
  \bibinfo {author} {\bibfnamefont {S.}~\bibnamefont {Nussinov}}, \ and\
  \bibinfo {author} {\bibfnamefont {Y.}~\bibnamefont {Zhang}},\ }\Doi
  {10.1016/j.physletb.2009.05.051} {\bibfield  {journal} {\bibinfo  {journal}
  {Phys. Lett.},\ }\textbf {\bibinfo {volume} {B677}},\ \bibinfo {pages} {311}
  (\bibinfo {year} {2009}{\natexlab{b}})},\ \Eprint
  {http://arxiv.org/abs/0903.2562} {arXiv:0903.2562 [hep-ph]} \BibitemShut
  {NoStop}%
\bibitem [{\citenamefont {Hamaguchi}\ \emph {et~al.}(2009)\citenamefont
  {Hamaguchi}, \citenamefont {Takahashi},\ and\ \citenamefont
  {Yanagida}}]{Hamaguchi:2009sz}%
  \BibitemOpen
  \bibfield  {author} {\bibinfo {author} {\bibfnamefont {K.}~\bibnamefont
  {Hamaguchi}}, \bibinfo {author} {\bibfnamefont {F.}~\bibnamefont
  {Takahashi}}, \ and\ \bibinfo {author} {\bibfnamefont {T.~T.}\ \bibnamefont
  {Yanagida}},\ }\Doi {10.1016/j.physletb.2009.04.070} {\bibfield  {journal}
  {\bibinfo  {journal} {Phys. Lett.},\ }\textbf {\bibinfo {volume} {B677}},\
  \bibinfo {pages} {59} (\bibinfo {year} {2009})},\ \Eprint
  {http://arxiv.org/abs/0901.2168} {arXiv:0901.2168 [hep-ph]} \BibitemShut
  {NoStop}%
\bibitem [{\citenamefont {Vertongen}\ and\ \citenamefont
  {Weniger}(2011)}]{Vertongen:2011mu}%
  \BibitemOpen
  \bibfield  {author} {\bibinfo {author} {\bibfnamefont {G.}~\bibnamefont
  {Vertongen}}\ and\ \bibinfo {author} {\bibfnamefont {C.}~\bibnamefont
  {Weniger}},\ }\href@noop {} { (\bibinfo {year} {2011})},\ \Eprint
  {http://arxiv.org/abs/1101.2610} {arXiv:1101.2610 [hep-ph]} \BibitemShut
  {NoStop}%
\bibitem [{\citenamefont {Choi}\ \emph {et~al.}(2010)\citenamefont {Choi},
  \citenamefont {Restrepo}, \citenamefont {Yaguna},\ and\ \citenamefont
  {Zapata}}]{Choi:2010jt}%
  \BibitemOpen
  \bibfield  {author} {\bibinfo {author} {\bibfnamefont {K.-Y.}\ \bibnamefont
  {Choi}}, \bibinfo {author} {\bibfnamefont {D.}~\bibnamefont {Restrepo}},
  \bibinfo {author} {\bibfnamefont {C.~E.}\ \bibnamefont {Yaguna}}, \ and\
  \bibinfo {author} {\bibfnamefont {O.}~\bibnamefont {Zapata}},\ }\Doi
  {10.1088/1475-7516/2010/10/033} {\bibfield  {journal} {\bibinfo  {journal}
  {JCAP},\ }\textbf {\bibinfo {volume} {1010}},\ \bibinfo {pages} {033}
  (\bibinfo {year} {2010})},\ \Eprint {http://arxiv.org/abs/1007.1728}
  {arXiv:1007.1728 [hep-ph]} \BibitemShut {NoStop}%
\bibitem [{\citenamefont {Bomark}\ \emph {et~al.}(2010)\citenamefont {Bomark},
  \citenamefont {Lola}, \citenamefont {Osland},\ and\ \citenamefont
  {Raklev}}]{Bomark:2009zm}%
  \BibitemOpen
  \bibfield  {author} {\bibinfo {author} {\bibfnamefont {N.~E.}\ \bibnamefont
  {Bomark}}, \bibinfo {author} {\bibfnamefont {S.}~\bibnamefont {Lola}},
  \bibinfo {author} {\bibfnamefont {P.}~\bibnamefont {Osland}}, \ and\ \bibinfo
  {author} {\bibfnamefont {A.~R.}\ \bibnamefont {Raklev}},\ }\Doi
  {10.1016/j.physletb.2010.02.050} {\bibfield  {journal} {\bibinfo  {journal}
  {Phys. Lett.},\ }\textbf {\bibinfo {volume} {B686}},\ \bibinfo {pages} {152}
  (\bibinfo {year} {2010})},\ \Eprint {http://arxiv.org/abs/0911.3376}
  {arXiv:0911.3376 [hep-ph]} \BibitemShut {NoStop}%
\bibitem [{\citenamefont {Buchmuller}\ \emph {et~al.}(2009)\citenamefont
  {Buchmuller}, \citenamefont {Ibarra}, \citenamefont {Shindou}, \citenamefont
  {Takayama},\ and\ \citenamefont {Tran}}]{Buchmuller:2009xv}%
  \BibitemOpen
  \bibfield  {author} {\bibinfo {author} {\bibfnamefont {W.}~\bibnamefont
  {Buchmuller}}, \bibinfo {author} {\bibfnamefont {A.}~\bibnamefont {Ibarra}},
  \bibinfo {author} {\bibfnamefont {T.}~\bibnamefont {Shindou}}, \bibinfo
  {author} {\bibfnamefont {F.}~\bibnamefont {Takayama}}, \ and\ \bibinfo
  {author} {\bibfnamefont {D.}~\bibnamefont {Tran}},\ }\Doi
  {10.1088/1475-7516/2009/09/021} {\bibfield  {journal} {\bibinfo  {journal}
  {JCAP},\ }\textbf {\bibinfo {volume} {0909}},\ \bibinfo {pages} {021}
  (\bibinfo {year} {2009})},\ \Eprint {http://arxiv.org/abs/0906.1187}
  {arXiv:0906.1187 [hep-ph]} \BibitemShut {NoStop}%
\bibitem [{\citenamefont {Adriani}\ \emph {et~al.}(2010)\citenamefont {Adriani}
  \emph {et~al.}}]{Adriani:2010rc}%
  \BibitemOpen
  \bibfield  {author} {\bibinfo {author} {\bibfnamefont {O.}~\bibnamefont
  {Adriani}} \emph {et~al.} (\bibinfo {collaboration} {PAMELA}),\ }\Doi
  {10.1103/PhysRevLett.105.121101} {\bibfield  {journal} {\bibinfo  {journal}
  {Phys. Rev. Lett.},\ }\textbf {\bibinfo {volume} {105}},\ \bibinfo {pages}
  {121101} (\bibinfo {year} {2010})},\ \Eprint {http://arxiv.org/abs/1007.0821}
  {arXiv:1007.0821 [astro-ph.HE]} \BibitemShut {NoStop}%
\end{thebibliography}%

\end{document}